**Transition region-induced kinetic Alfven wave conversion. Electric displacement field**

Petko Nenovski

Independent researcher, Sofia, Bulgaria

Abstract

The transition region is a thin inhomogeneous region where Alfven waves' energy fluxes generated elsewhere in the solar atmosphere are effectively converted. Large-scale kinetic Alfven wave propagations, transmission, and reflection processes across the transition region are examined. The two-fluid model is adopted, and a study is conducted of how the kinetic Alfven wave electric displacement field changes across the transition region. The analysis outcomes are: the kinetic Alfven wave and electrostatic ion sound waves are coupled by the transverse wavenumber; wave electric field components (normal to the transition region) become enhanced up to two orders; the energy fluxes of transmitted kinetic Alfven waves are re-directed almost horizontally along the transition region, an evanescent electric field zone of enhanced intensity is induced tightly beyond the transition region; the ponderomotive force that emerges in that zone due to the reflected kinetic Alfven waves accelerates plasma particle upwards compared to their initial energy in the upper chromosphere.

## Introduction

Alfven waves and their possible interactions with plasma particles at various spatial scales (ranging from Debye/particle cyclotron lengths to those comparable with the plasma density/magnetic fields structures as a whole) are critical in space, astrophysical, and laboratory plasmas resulting in particle acceleration, wave mode conversion, energy and mass transport, plasma heating, etc (Jess at al., 2009; 2020; Mathioudakis et al., 2012; Liu et al., 2019; Wygant et al., 2000). High-resolution observations have clearly demonstrated the ubiquitous presence of Alfven waves in the solar atmosphere (Morton et al. 2012; 2023).

Oscillations with periodicities ranging from 126 to 700 seconds originating above the bright points have been revealed. These oscillations are a signature of Alfven waves produced by a torsional twist (Jess et al., 2009). Observational evidence of a chromospheric resonance cavity existing above a magnetic sunspot has been provided (Jess et al., 2020). MHD waves with phase speeds of 1 to $4 \times 10^3$ km/s and trajectories consistent with the direction of the magnetic field have been recorded (Tomszyk et al., 2007). Crammer et al (2007) have observed and estimated the possible energy fluxes of the Alven wave modes upward and downward in the chromosphere to be of about 5 (4 to 7) kW/m$^2$. Chromospheric swirls in the transition region and low corona have been identified as observational signatures of rapidly rotating magnetic structures, being proposed as an alternative for transferring energy from the lower into the upper solar atmosphere region (Wedemeyer-Böhm et al., 2012).

Alfven wave-plasma interactions are well manifested at plasma density steep gradients that exist in space plasmas. Density gradients are often considered as a boundary structure between plasmas with different populations and content, either the magnetopause (between solar wind and Earth's magnetosphere), or the plasma sheet boundary layer (PSBL) in the magnetosphere. Similar steep plasma density gradients steadily exist in the solar atmosphere on global or local scales. Kinetic Alfven waves carrying a parallel electric field along the background magnetic field have been proposed (Hasegawa, 19756; Hasegawa and Chen, 1974,1976; Hasegawa and Mima, 1978) and suggested as an effective mechanism for wave dissipation (Lysak 2023; 2007; Chen et al., 2021), particle acceleration (Wygant et al., 2000). Kinetic and ideal Alfven wave conversion of different scales is expected in the solar transition region (TR) – a thin, highly dynamic, and inhomogeneous layer of the Sun's atmosphere located between the cool chromosphere and the hot corona (Cally, 2022)

Recent observations from the Daniel K. Inouye Solar Telescope (DKIST) have, for the first time, revealed evidence of electric fields at magnetic diffusion sites in the solar chromosphere. Anan et al (2024) have detected the polarization signature of electric fields associated with magnetic diffusion. They have measured the linear and circular polarization across the hydrogen Hε Balmer line at 397 nm at the site of a brightening event in the solar chromosphere. Based on these observations, a nearly horizontal electric field does exist at a null point of the observed large-scale magnetic field structure, consistent with the magnetic reconnection in the antiparallel magnetic field structure (Anan et al, 2024). Under the assumption of a typical Alfvén velocity of 100 km $s^{-1}$ and a strength of the reconnecting magnetic fields of 0.05 T, the horizontal electric field would amount to 100 V/m there.

To date, spatially resolved Alfven waves carry energy flux and potentially serve as an energy reservoir sufficient to heat the solar corona (McIntosh et al., 2011; Samanta et al., 2019; Soler, 2025). Solar corona heating mechanisms are not fully understood. Irrespectively of the multitude of observational facts about the enormous Alfven wave energy fluxes initiated during reconnections of various scales, the problems of wave energy transfer and how and where is dissipated are still debated.

In this connection, processes of Alfven wave propagation, transmission and reflection, mode conversion at plasma-plasma boundaries have been well studied for a long time, both in space and astrophysical aspects (Mathioudakis et al., 2012; Chae and Lee, 2019; Piantschitsch et al., 2020; Cally, 2022; Murabito et al., 2024a,b; Kumar et al., 2025). The mode conversion of slow, Alfvén and fast magnetohydrodynamic waves injected at the base of a two-isothermal-layer stratified atmosphere with a uniform magnetic field, crudely representing the solar chromosphere and corona with intervening discontinuous transition region, has been examined by Cally (2022). Based on the results found there, all three injected wave types can partially transmit as coronal Alfvén waves in varying proportions dependent on frequency, magnetic field inclination, wave orientation, and distance between the Alfvén/acoustic equipartition level and the transition region. The conclusion drawn there is that net Alfven wave transmission to the corona is limited, and additional magnetic field structuring probably may be required to provide sufficient wave energy flux.

Mechanisms of MHD wave mode conversion into localized kinetic Alfven wave within the plasma boundaries have been studied (Hasegawa and Chen, 1976; Hasegawa, 1977); Observations (Wigant et al, 2000) suggest that the small-scale kinetic Alfven waves are generated from the larger-scale Alfven waves. The same observations provide strong evidence that in addition to the auroral particle energization processes occurring at altitudes between 0.5 and 2 Earth's radii, important heating and acceleration mechanisms are operating at these higher altitudes within the plasma sheet. Wygant et al.'s in situ observations and analyses have shown that parallel and perpendicular electric fields of KAW in the Earth's magnetosphere are of the order ~ 1 mV/m and ~ 0.3 V/m, respectively. Alfven wave conversion into kinetic Alfvén Waves (KAWs) is proposed in the solar atmosphere. KAWs can effectively trigger energy channeling from fluid scales to kinetic scales, causing plasma heating and accelerating ions and electrons (De Pontieu et al., 2007). KAWs are thus well studied in the context of magnetospheric and auroral physics (Lysak, 2023). At propagation oblique to the background magnetic field $\mathbf{B}_o$, the damping rate of KAW modes scales as $k_\perp^2 k_\parallel$ where the subscripts denote directions relative to $\mathbf{B}_o$. The authors have found that this damping "progressively (although not monotonically) increases with increasing electron and proton $\beta$, corresponding to four distinct damping regimes: non-resonant, electron Landau, proton Landau, and proton transit-time damping (Gary and Borowski, 2008). A detailed and extensive study focused on kinetic Alfven waves with small spatial scales to examine the perturbed electromagnetic (EM) fields, the Poynting flux, net power transfer through the solar flux loop tubes, resonant particles' speed, group speed, and the damping length of KAWs has been performed in depth by Ayaz et al (2025) including arbitrary electron-to-ion temperature ratios (Te/Ti) and suprathermal particles in the solar corona (Ayaz et al, 2025).

Kinetic Alfven waves have a parallel electric field and are a highly damped mode, especially at small scales. The plasma inhomogeneities do enforce various small-scale kinetic and large-scale fluid-like effects. The solar transition region (TR) is a remarkable plasma-plasma boundary. It lies between the more dense and cool (T ~ 1 eV) weakly ionized chromospheric plasma and the hot (T ~100 eV) and low-density corona. On the one side, Alfvén wave fluxes upward from the solar photosphere are effectively reflected in the solar transition region due to strong density gradients there. So, the transition region acts as a barrier for very intense Alfven waves of various kinds (shear, kink, etc) in the chromosphere, reflecting them downward, where they are ultimately dissipated. On the other hand, in the TR, neutrals can be ionized, and particles can be accelerated/decelerated. Radio bursts are expected to originate there. It can be supposed that investigations of the Alfven wave-TR system itself might be critical to studying the corona heating mechanisms, the particle acceleration, and the origin of the solar wind, as well as the initiation mechanisms of solar flares and coronal mass ejections (CMEs), and so on.

Standing Alfvén waves (e.g., in the chromosphere, spicules, coronal loops) of enhanced amplitude may create a time-averaged ponderomotive force (Hollweg, 1876; Allan, 1993). These ponderomotive forces are due to the gradient in the nonlinear magnetic pressure associated with the wave's reflection acting as a force upward on the plasma ions in the chromosphere region. Laming (2017), Murabito et al (2024) have investigated the 'chromospheric' Alfvén waves in detail, which give rise to the separation of ions and neutrals - the so-called first ionization potential (FIP) effect - through the action of the ponderomotive force of those Alfvén waves.

The present study focuses on the rate of kinetic Alfven wave energy conversion at and across the transition region (TR). Large-scale effects occurring upon traversing the solar transition region are only considered. The adjacent upper chromosphere and the base of the corona are modeled as follows: a uniform plasma of certain densities and temperatures, both immersed are in a background magnetic field oblique to the normal of the transition region. Plasma pressures throughout the transition region boundary are balanced, allowing, within the framework of two-fluid hydrodynamics and the full set of Maxwell equations, examination of fluid motion, field transmission, and the reflection of kinetic Alfven wave modes. Reflected Alfven waves enforce an evanescent field zone where ponderomotive forces arise. Waves' electric field enhancement across the TR is addressed and quantified.

Within an approach, the large-scale Alfven wave energy interaction processes in the Chromosphere-Transition region-Corona system and associated boundary effects on Alfvén waves propagation across the boundary of two adjacent plasma media – chromosphere and corona being in contact via the solar transition region (TR), are considered. The Alfven wave parallel and perpendicular electric fields change magnitudes by several orders of magnitude. Associated ponderomotive forces are also examined.

## 1.Kinetic Alfven and Ion Sound waves coupled by transverse wavevector $\boldsymbol{k}_{\perp}$

An extended and thorough consideration of low-frequency shear Alfvén waves should include both kinetic and inertial Alfven waves. This suggests the two-fluid hydrodynamic and/or Hall magnetohydrodynamic equations and a full set of Maxwell equations (Wu and Fang, 1999; Yang et al., 2014, Mallet et al., 2023).

### 1.1. Dispersion equation of KAW and ISW

Assuming a frame of reference with a uniform magnetic field $\boldsymbol{B}_0$, $(B_0 = (B_0 \sin(\theta), 0, B_0 \cos(\theta))$, and a linear wave electric field $\boldsymbol{E}$ lying in the O$xz$ plane, $\boldsymbol{E} = (E_x, 0, E_z)$, (see Figure 1), we are looking for solutions in the form

$$\sim A exp(ik_x x + ik_z z - i\omega t).$$

Introducing notations $\varepsilon_\perp$ and $\varepsilon_\parallel$ denoting the perpendicular and parallel dielectric functions of the plasma in the low-frequency (MHD along and perpendicular to $\boldsymbol{B}_0$ ) limit ($\omega \ll \omega_{ci}, \omega_{ci}$ is the ion cyclotron frequency), it is possible to derive the following dispersion equation of kinetic Alfven waves:

$$\varepsilon_\perp \varepsilon_\parallel - \varepsilon_\parallel {n_\parallel}^2 - \varepsilon_\perp {n_\perp}^2 = 0, \quad (1)$$

where $k_\perp$ and $k_\parallel$ are the perpendicular and parallel wavenumbers, respectively, where

$${n_{\parallel,\perp}}^2 \equiv \frac{c^2 k_{\parallel,\perp}^2}{\omega^2}, \quad (2)$$

$$\varepsilon_\perp = 1 + \frac{c^2}{v_A^2}, \quad (3)$$

$$\varepsilon_\parallel = 1 - \frac{\omega_{pi}^2}{\omega^2} + \frac{1}{k_\parallel^2 \lambda_D^2}, \quad (4)$$

$n_{\parallel,\perp}$ are the refractive indices, $\varepsilon_\perp$, and $\varepsilon_\parallel$ - the perpendicular and parallel dielectric functionss derived under the following constraints: the wave frequency, ω, is less than both the ion cyclotron frequency $\omega_{ci}$ and the ion plasma frequency $\omega_{pi}$, also $k_\parallel v_{Ti} \ll \omega \ll k_\parallel v_{Te}$ , $v_{Ti,e}$ – thermal velocity of ions and electrons. $v_A$ is the Alfven velocity; $\lambda_D \equiv \frac{v_{Te}}{\omega_{pe}}$ - the electron Debye length, $c$ - the light velocity.

Ignoring the perpendicular wavenumber $k_\perp$, ($k_\perp \to 0$), the dispersion equation reduces to

$$\varepsilon_{\parallel}(\varepsilon_{\perp} - n_{\parallel}{}^{2}) = 0, \tag{5}$$

which results in two independent (low-frequency) wave modes: shear Alfven wave $\omega_A$, and ion sound $\omega_S$ waves:

$$\omega_A = k_{\parallel} v_A, \; (v_A \equiv \frac{B_0}{\sqrt{\mu_0 \rho_0}}), \tag{6}$$

and

$$\omega_S \cong k_{\parallel} v_S, \; (v_S \equiv \sqrt{\gamma \frac{kT_e}{m_i}}), \tag{7}$$

The second wave branch (7), the ion sound mode, appears under the condition that the frequency ω should be much less than the ion plasma frequency $\omega_{pi}$ ($\omega \ll \omega_{pi}$)

Under the condition $k_{\perp} \neq 0$, both modes shear Alfven waves, and the ion acoustic waves, however, are no longer independent of each other. Shear Alfven waves, acquiring a perpendicular wavelength $k_{\perp}$ of any scale, becoming kinetic Alfven waves are necessarily accompanied by ion acoustic waves of the same scale. The two wave modes are necessarily coupled even in uniform, homogeneous plasmas.

Under $k_{\perp} \neq 0$ and $\omega \ll \omega_{pi}$ , $k_{\parallel} v_{Ti} \ll \omega \ll k_{\parallel} v_{Te}$ dispersion equation (1) reads

$$(\varepsilon_{\perp}\omega^2 - c^2 k_{\parallel}^2)(\, 1 - \frac{\omega_{pi}^2}{\omega^2} + \frac{1}{k_{\parallel}^2 \lambda_D^2}) - \varepsilon_{\perp} c^2 k_{\perp}^2 = 0. \tag{8}$$

Approximating the perpendicular permittivity $\varepsilon_{\perp} = 1 + \frac{c^2}{v_A^2} \cong \frac{c^2}{v_A^2}$, the dispersion equation yields the following spectra:

$$\omega_{KAW}^2 \cong k_{\parallel}{}^2 v_A{}^2 (1 + k_x^2 \rho_S^2), \tag{9}$$

and

$$\omega_{ISW}^2 \cong k_{\parallel}^2 v_S^2 (1 - k_{\perp}^2 d_i^2), \tag{10}$$

where $\rho_s = \frac{v_S}{\omega_{ci}}$ is the ion sound cyclotron radius, and $d_i = \frac{c}{\omega_{ci}}$ is the ion inertial depth. The first spectrum is consistent with the kinetic Alfven wave dispersion under the fluid conditions mentioned above (Gary and Borowsky; 2008 Fayad et al., 2023; Lysak, 2023, and references therein), while the second one (10) should be consistent with the low-frequency electrostatic ion sound mode spectrum in magnetized plasmas (Basu et al., 1985).

Co-existence of both the kinetic Alfven wave and the ion sound wave enforces a joint examination of the propagation, transmission, and refraction of these modes across the steep boundary of plasmas of different properties. The transition region (TR) in the solar atmosphere represents such a boundary. Further, we focus on the solar transition region and address possible changes in Alfven wave characteristics and in Alfven wave energy conversion there.

**1.2.Electric displacement field $\boldsymbol{D}$ at the transition region (TR).**

The boundary conditions for electromagnetic field propagation and transmission across an interface with no free surface charge enforce the normal component of the electric displacement field $\boldsymbol{D}$ ( $\boldsymbol{D} = \hat{\varepsilon}.\boldsymbol{E}$, where $\hat{\varepsilon}$ is the dielectric permittivity tensor) to be continuous.

$$\boldsymbol{D}_{n,1} = \boldsymbol{D}_{n,2}. \qquad (11)$$

This rule should be preserved for the boundaries of magnetoactive plasmas, regardless of the orientation of the ambient magnetic field $\boldsymbol{B}_0$ with respect to that boundary. The solar transition region TR, a very thin region of approximately hundreds of km lying horizontally over the chromosphere region, represents such a plasma boundary. The upper chromosphere and the base of the corona, being permeated by strong magnetic fields of arbitrary orientation, are detached by this transition region boundary.

The boundary conditions for the magnetic fields also require continuity of the normal component $\boldsymbol{B}_n$, while the tangential component is determined by the surface currents flowing within the boundary plasma structure.

The impact of the displacement field $\boldsymbol{D} = \hat{\varepsilon}.\boldsymbol{E}$, parallel to the constant magnetic field $\boldsymbol{B}_0$, manifests itself at large scales when a steep boundary is formed between plasmas in different conditions (e.g., in low- or high ionization states) and/or different basic parameters (density, temperature, etc).

Having chosen a reference frame with axes z oriented along the normal of the boundary (fig. 1 ), the continuity condition for the electric displacement field $\boldsymbol{D}$ across the boundary ($z = 0$) results in the following equation

$$D_{z,1} = D_{z,2} \tag{12}$$

where

$$D_z = D_{\parallel}\cos(\theta) + D_{\perp}\sin(\theta) \tag{13}$$

with $D_{\parallel} = \varepsilon_{\parallel}E_{\parallel}$ , $D_{\perp} = \varepsilon_{\perp}E_{\perp}$ , and approximate expressions of $\varepsilon_{\parallel}$ and $\varepsilon_{\perp}$:

$$\varepsilon_{\parallel} = 1 - \frac{\omega_{pi}^2}{\omega^2} + \frac{1}{k_{\parallel}^2\lambda_{De}^2} \cong -\frac{\omega_{pi}^2}{\omega^2} + \frac{1}{k_{\parallel}^2\lambda_{De}^2}, \varepsilon_{\perp} = 1 + \frac{c^2}{v_A^2} \cong \frac{c^2}{v_A^2}. \tag{14}$$

the parallel (along the magnetic field $\boldsymbol{B}_0$ and perpendicular permittivities of plasmas under the low frequency condition ($\omega \ll \omega_{ci}$); $k_{\parallel}$ is the parallel wavenumber, $\lambda_{De}$- the Debye length, $\omega_{pi}$ -the ion plasma frequency, $v_A$ - the Alfven velocity, c – the light speed, and ω -the wave frequency.

Given the approximation expressions of both the dielectric functions $\varepsilon_{\parallel}$ and $\varepsilon_{\perp}$, equation (12) suggests the following relations about both the electric fields $E_{\parallel}$ and $E_{\perp}$ across the boundary z =0:

$$E_{\parallel,2} = \frac{\varepsilon_1}{\varepsilon_2}E_{\parallel,1}, \text{ and } E_{\perp,2} = \frac{\varepsilon_1}{\varepsilon_2}E_{\perp,1}. \tag{15}$$

The ratio of the permittivity tensor components, however, is proportional to the ratio of the plasma densities:

$$\frac{N_{0,1}}{N_{0,2}} \gg 1 \tag{16}$$

valid under the following conditions

$$\omega \ll \omega_{pi}, v_A \ll c.$$

**1.3.Electric displacement field induced effects**

The results obtained above should be valid in the approximation that the spatial scales of the Alfvén waves exceed the finite thickness $d_{TR}$ of the transition region TP.

In this approximation, the magnitude of the electric field increases with the ratio of the plasma densities in the upper chromosphere and the base of the corona. It is about two orders of magnitude. The following significant effects are identified.

First, the equipartition of kinetic and magnetic energy, which is inherent in shear and in ideal MHD, is not preserved. The dissipative losses of kinetic Alfvén waves governed by the Landau damping would increase considerably. These losses arise from and are at the expense of the magnetic energy of Alfvén waves, which are effectively transformed into TR.

The energy equipartition violation occurs at an arbitrary orientation of the external magnetic field $\boldsymbol{B}_0$ relative to the normal of the TP. In the case that the magnetic field $\boldsymbol{B}_0$ is horizontal, i.e. coplanar to the TR boundary surface, the condition for continuity of the displacement field $\boldsymbol{D}$ across the TP boundary is fully determined by the transverse magnetic field part of the permittivity tensor: $\varepsilon_{\perp}$ - inversely proportional to the squared Alfvén velocity $v_A^2$, i.e. the increase in the electric field normal to the boundary again reduces to the ratio of the plasma densities on both sides of the boundary $z$ = 0 (16): $\frac{N_{0,1}}{N_{0,2}} \gg 1$.

This conversion is consistent with the conservation of the wave energy flux through the transition region: as the plasma density decreases, the velocity amplitude, respectively the electric field, increases quadratically.

The condition for the conservation of energy density, the constant wave frequency ω, and the condition for the continuity of the wave vector $k_{\perp}$ enforce a condition that the wave number $k_{\|}$ varies as ${v_A}^{-1} \sim \sqrt{N}$. The transmitted Alfven wave energy flux is enforced to propagate (nearly horizontally) along the transition region because the vertical wavelength $k_{z,2}$ (in region 2) becomes considerably less than the Alfven wavelength $k_{z,1}$in the upper chromosphere (see Appendix A).

On the other hand, the ratio of the two components of the electric field, given by the formula (15), implies that the transverse electric field $E_{\perp}$ in region 2 varies to $\sqrt{N}$, which is compatible with the condition for the continuity of the displacement field (12). The magnetic component of the Alfvén wave field then varies proportionally to the plasma density $N$:

$$b_y = k_{\|} \frac{E_{\perp}}{\omega} \sim \sqrt{N}.\sqrt{N} \sim N$$

In other words, in region 2 (the base of the corona) the magnetic energy of the Alfvén wave decreases in direct proportion to the decrease in the plasma density.

According to the conditions for Alfvén wave reflection (Piantschitsch, 2020; Cally, 2022), the transmission and reflection coefficients of mass velocities were derived from linear theory to calculate estimations for phase speeds of incoming, reflected, and transmitted waves. The fraction of the Alfvén wave energy passed into region 2 is thus proportional to

$$\sim\left(\frac{v_{A_1}}{v_{A_2}}\right)\sim\frac{N_2}{N_1}.$$

Given the ratio between the electric and magnetic parts of the energy density of the energy passed into region 2, we find that the energy of the kinetic Alfvén wave passed is practically transformed into electrostatic energy. The physical meaning of this result can be interpreted by assuming that the full kinetic energy of the kinetic Alfvén wave incident on the TP passes practically unhindered into region 2 (into the base of the corona). This conclusion is consistent with the notion that the boundary (TP) must be free (as opposed to the rigidly fixed boundary, where the displacement (field, velocity) is zeroed), which means that the motion of the particles is free. Hence, when passing through the boundary, the particles (electrons and ions) retain their momentum and energy.

The energy of the kinetic Alfvén wave that has passed through the boundary (TP) should be deposited in region 2, as a result of its cumulative conversion across the TP. Since the kinetic Alfvén wave is dispersive, and the ratio of the longitudinal to transverse wavenumber is usually very small ( $k_{\parallel}/k_{\perp} << 1$, its group velocity in the direction of $k_{\perp}$ is dominant:

$$\frac{v_{gr}}{v_{ph}} \cong k_{\parallel}k_{\perp}\rho_s^2$$

It follows that its energy flux propagates almost horizontally along the boundary (TP), where it should ultimately be completely absorbed.

One may conclude that during the reflection process from the transition region TP, the incident Alfven wave practically loses its kinetic energy associated with the electrostatic field and returns to the chromosphere weakened (without the kinetic energy converted into the electrostatic field $E_z$).

## 2. Estimation of the KAW and ISW evanescent field zones

Alfven wave field beyond the plasma-vacuum boundary, where the wave is reflected, evanesces in an exponential form

$$\sim exp(-\kappa z)$$

where the coefficient $\kappa$ characterizes the field penetration spatial scale. At the plasma-plasma boundary, the field penetration depth depends on the polarization and/or conductivity properties of the region beyond the reflection point. In a region characterized by real conductivity $\sigma$, the penetration depth should be determined by a skin depth $\delta$ equal to $\frac{1}{\sqrt{\sigma_0 \omega \sigma}}$.

The skin depths (classical and anomalous, etc) are widely studied in the literature. Presumably, the kinetic Alfven waves of scales comparable to the ion gyroradius have a high Landau damping rate. For them, the penetration depth is determined by the Landau damping rate $\gamma$ ($\gamma \sim \sigma^{-1}$).

For large-scale kinetic Alfven waves the Landau damping effects, however, are negligible. Instead, large-scale kinetic Alfven waves are (partly) transmitted and (partly) reflected upon falling on the transition region (TR). The field penetration depth, then, should be conditioned on the requirement that the wave reflection process be executed. This suggests a relevant plasma motion in region 2 in a zone of some size (attached to the TR boundary). The depth of that zone has to be determined by the actual wave dispersion relation in the (dissipative or non-dissipative) region in question. Let us examine the case in detail.

### 2.1. Alfven wave evanescent field

The energy density flux of large-scale shear Alfven waves incident on the plasma-plasma boundary should be reflected according to Snell's law. The incident Alfven wave travelling from the photosphere heights meets a rarefied plasma of much lower plasma density (region 2, the solar corona). Accordingly, the reflected Alfven wave's magnetic field perpendicular to the boundary (z = 0) is almost cancelled. The associated electric field of the reflected Alfvén wave should then be added to that of the incident wave, depending on the ratio of the Alfven velocities on both sides of that boundary (Cally, 2022). For ideal shear Alfven waves, the parallel magnetic field component, by definition, is zero.

Upon Alfven wave reflection from a rarefied plasma medium ($\frac{v_{A_1}}{v_{A_2}} \to 0$): $E_x^{in} \cong E_x^r,\ \ b_x^{in} \cong -b_x^r$ where $k_x \ = k_x^r$ and $k_z^{in} \ = \ -k_z^r$. At the boundary surface, the wave magnetic field $b_y$ becomes nearly zero. In region 2, the wave magnetic field $b_y$ is practically zero, implying

$$rot\ E \cong 0,$$

In the absence of the wave magnetic field, and having in mind that the Alfven wave electric field $\boldsymbol{E}$ is almost oriented along the wavevector $\boldsymbol{k}_x$, the total electric field $\boldsymbol{E}$ (perpendicular and transverse ones) behaves in the rarefied plasma region 2 as a quasi-electrostatic disturbance. (Upon reflection, the wave's magnetic field returns to the chromosphere).

As for the dynamical boundary conditions, the mass velocity $v$ continuity across the boundary reads

$$div\ v_{\perp} = 0. \tag{17}$$

Eq. (17) enforces a wave evanescent solution of the form

$$\sim A \exp(-\kappa z) \exp\left(ik_x x - i\omega t\right), \tag{18}$$

where the attenuation coefficient $\kappa$ is determined by the relation

$$\kappa \ = k_x. \tag{19}$$

The same attenuation characteristic holds for the transverse electric field as far as the perpendicular electric field: $E_{\perp} \ = -v{\times}B_0$ is concerned. It is not valid for the more general case of plasma motion along the background magnetic field $\boldsymbol{B}_0$ involved by kinetic Alfven waves. The parallel electric field should be taken into account.

2.2. Ion sound wave evanescent field

The kinetic Alfven wave and the associated ion sound waves support parallel electric fields. Presumably, the ion sound wave field $E_{\parallel}$ should attenuate beyond the kinetic Alfven wave reflection point. Looking for a solution in the form (18), beyond the reflection point z=0, the Poisson equation of $E_{\parallel}$ reads:

$$\varepsilon_i \Delta E_\parallel + \frac{\omega_{pe}^2}{v_{Te}^2} E_\parallel = [-\varepsilon_i (k_x^2 - \kappa^2) + \frac{\omega_{pe}^2}{v_{Te}^2}] E_\parallel = 0. \quad (20)$$

Where the ion dielectric permittivity $\varepsilon_i$ is given by expressions (3,4). The non-trivial solution of (20) ensures

$$\kappa^2 = k_x^2 + \frac{1}{\varepsilon_i} \frac{\omega_{pe}^2}{v_{Te}^2}$$

which, in the long wavelength limit $\varepsilon_i \cong -\frac{\omega_{pi}^2}{\omega^2}$, reduces to

$$\kappa^2 = k_x^2 - \frac{\omega^2}{v_S^2} \cong k_x^2 \; . \quad (21)$$

The perpendicular electric fields $E_\perp$ of the kinetic Alfven waves should behave in the same way – it exponentially attenuates with the same attenuation coefficient $\kappa$.

The found attenuation coefficient $\kappa$ for the waves: kinetic Alfven and associated ion sound waves is consistent with the wave reflection mechanism requiring plasma motion and volume of a certain depth necessary to enforce magnetic field flux backward to region 1. This happens by enforcing the ion plasma inertial-motion response in the field evanescent zone. As a result, the depth is found to be inversely proportional to the wavenumber parallel to the transition region boundary. The large-scale model requirement says that the evanescent field depth of penetration found should exceed the transition region thickness $d_{TR}$.

As for a vacuum ($v_S^2, v_{A2} \rightarrow 0$) , as expected coefficient $\kappa_2$ does coincide with $k_x$.

In essence, the wave field, the Alfven wave electric field $E_\perp$ beyond the boundary, should be considered rather as an electrostatic-like one than an electrostatic one. The Alfven wave magnetic field $b_\perp$ in region 2 cannot be fully neglected also for the following reasons: i) the wave field penetrates magnetized plasma, and ii) the wave reflection mechanism itself is not perfect – a small part of the incident wave still transmits across the boundary. If medium 2 of another plasma structure (of different plasma density, temperature, etc.), the incident Alfven wave is partially transmitted, depending on the Alfven velocity ratio

$$\frac{v_{A2}}{v_{A1}}.$$

It is worthing to say this result is derived in the non-dissipative limit, where collisions and Landau damping (effective in the short wavelengths limit) are not taken into account. The following considerations reassume the possible effects connected with the mentioned dissipation mechanisms.

In most cases, the field penetration depth corresponds to the skin effect that occurs when electromagnetic waves penetrate a conducting medium of certain real conductivity. The thickness of the skin effect decreases with increasing conductivity. In a plasma with dielectric properties, the penetration depth of the wave field is determined by the wave frequency and the wave dispersion in the corresponding medium. If the electric field $E_{\parallel}$ of the kinetic Alfvén waves, was to be defined by the magnitude of the Landau damping, which is significant, the penetration depth of the field $E_{\parallel}$ would have to be practically zero. Fortunately, in the case of large-scale kinetic Alfven waves, the Landau damping is of negligible magnitude and thus neglected. Besides, the collision frequency in the transition region and the base of the corona remains low. Hence, region 2 could be specified as a non-dissipative medium. The mechanism of Alfvén waves reflection from the second medium would then occur.

Both the kinetic Alfven waves and the associated ion sound waves evanesce beyond the transition region, having attenuation coefficients close to the transverse wavenumber $k_x$ (Figure 2).

### 3.1. Alfven wave reflection-induced ponderomotive force $F_p$

The electric field enhancement of large-scale kinetic Alfvén waves across the chromosphere reasonably raises the question of nonlinear effects associated with their transmission and reflection at the TR. The ponderomotive force of Alfvén waves manifests itself in the course of their propagation upward to the TP in the presence of the plasma density stratification in height (Holweg, 1978; Laming et al., 2017). A study of non-linear wave effects and the ponderomotive force, in particular, becomes urgent.

In part 2 the exponential decrease of the kinetic Alfvén wave field amplitude is quantified and estimated. It is located behind the TR reflection point, where the ponderomotive force should be specified. The ponderomotive force for the kinetic part of the shear Alfven waves reads (Young and Gourdain, 2025)

$$F_{p,AW} = -\frac{e^2}{4m_e\omega^2}\,\nabla(|E_\perp|^2), \quad (22)$$

Th The ion sound wave electric field, initiated by large-scale kinetic Alfven wave penetration in the TR, also becomes enhanced in region 2 (the base of the corona). The ponderomotive force, $F_{p,IA}$, is conventionally expressed as a result of the spatial gradient of the time-averaged parallel electric field $E_\parallel$:

$$F_{p,ISW} = -\frac{e^2}{4m_e\omega^2}\,\nabla(|E_\parallel|^2), \quad (23)$$

Both ponderomotive forces act on the plasma particles starting at the transition region and accelerating them upward into the lower corona (Figure 2). Forces push electrons out of the high-intensity zone of the evanescent fields. The exponential form of the electric fields beyond the reflection boundary results in the following expression ():

$$F_p = -\frac{e^2}{4\omega^2}\nabla(|E_\perp|^2) \sim \frac{e^2}{2\omega^2}\,\kappa E_\perp^2, \quad (24)$$

suggesting that the estimated Alfven wave reflection-induced ponderomotive force $F_p$ (22,23) is directed upward and placed closely beyond the TR.

### 3.2. Ponderomotive force $F_p$ estimation. Particle energy gain.

The kinetic energy of the reflected Alfvén wave penetrates to a height beyond the transition region (TR). This is the region where the ponderomotive force $F_P$ is maximum and acts to accelerate electrons and ions in the same direction – upwards. The penetration depth found is of the order of the reciprocal of the wavenumber $k_\perp = \frac{2\pi}{\lambda_\perp}$. An estimate of the upper limit of the energy that ionized particles can acquire is evaluated. This value follows from the quantitative estimate of the work performed by the ponderomotive force (24) of the Alfvén wave due to the electric field $E_\perp$ penetrating beyond the boundary:

$$\mathcal{E}_{kin} \cong F_p \lambda_\perp,$$

where $\mathcal{E}_{kin}$ is the kinetic energy of the accelerated particles, and $\lambda_\perp$ is the height scale of the region where the reflected Alfvén wave field penetrates. Let us estimate these quantities quantitatively. The scale $\lambda_\perp$ is of the order of $10^2$ km and less. We assume that the component $E_\perp$ can be inferred from the experimentally observed oscillation velocity of observed plasma structures (e.g. spicules in the chromosphere). It is of the order of 5-10 km/s. With an average value of the magnetic field $B_0$ in the chromosphere of about 1-10 gauss, we find that $E_\perp$can vary between 0.5 and 10 V/m. $E_\perp$magnitude rather close to the lower limit is chosen: $E_\perp$ is 0,5 V/m. The field $E_\perp$ magnitude should increase by roughly two orders, depending on the ratio $\frac{N_1}{N_2}$.

Under these assumptions, the estimate of the work (energy) performed by the ponderomotive force $F_p$ in the zone of the wave electric field penetrating above the TP with a size $d_{TR}$ of about $10^2$ km is

$$\mathcal{E}_{kin} \cong 10^2 \text{ eV}$$

This estimate is roughly an upper bound due to the assumption that the ponderomotive force $F_p$ is constant. In reality, the process is dynamical and determined by the upward flow of Alfvén wave energy, as well as by the wave reflection rate at the TP.

Such particle acceleration occurs (via the directed upward ponderomotive force) in the zone of exponentially decaying electric field above the TR, i.e., the acceleration of particles (electrons and ions) and the energy gain should be deposited in the solar corona above the TR.

The main characteristics of the ponderomotive force $F_p$, defined above, are:

- It is a result of the Alfvén wave field amplitude inhomogeneity itself, and is placed beyond the wave reflection point at TR;
- The ponderomotive force is rooted in the kinetic part of the energy of the reflected Alfvén waves. That part of energy is lost above the reflection point, i.e., in and definitely above the TR. The wave magnetic energy practically returns to the chromosphere, where it is ultimately lost;
- The mechanism of acceleration of charged particles works beyond the reflection point, above the transition region TR;
- The acceleration of electrons and ions can be significant and should grow up to the thermal energy of the plasma in the corona, being of the order of 100 eV. This Alfven wave reflection-induced acceleration mechanism arises when Alfven waves impact TR from below.

The difference in plasma temperatures and densities between the two media – the upper chromosphere and the base of the corona – is a fact. This circumstance reflects the magnitude of the ratio of the corresponding dielectric functions, which is usually of the order of 2 to 3 orders of magnitude. Accordingly, the electric field components increase by the same factor. It can be assumed that a further increase in this factor would lead to an even greater increase in the electric field, accompanied by higher ionization in the high chromosphere, penetration of ionized particles through the TP, and an increase in the plasma density in the base of the corona. Observationally, such a scenario is missing. The associated pressure balance across the TR remains in steady-state balance, the ratio of plasma densities and temperatures persists over time.

**4. Discussion**

Let us outline the applicability of the results obtained during the analysis.
First, any plasma-plasma boundary encompasses a region of plasma density gradients that generate various plasma instabilities with scales comparable to the plasma parameter gradients. In the TR, the plasma density drops steeply by 2-3 orders. The plasma temperature increases across the TR by the same order. Overall, the plasma kinetic pressure remains 'steady-state' throughout the TR. These peculiarities (strong gradients over short distances) suggest an integrated approach, conventional in the MHD surface wave theory: TR to be considered as an infinitely thin boundary, allowing both the regions (region 1 and region 2) to be involved and treated as adjacent plasma-plasma structures at that boundary, considered as a system. This 'surface wave-like' approach requires adequate conditions on the spatial scales of the large-scale Alfven waves under consideration, namely:

$$k_n d_{TR} < 1.$$

where scale $k_n^{-1}$, (n denotes the wavenumber normal to the boundary) should exceed the TR thickness $d_{TR}$. Such an integrated approach enlightens the Alfven wave reflection mechanism at the TR boundary. The Alfven wave field encountering TR is partly reflected back into region 1 from where the wave arrives, and partly transmitted into region 2.

Second, when the Alfven wave field is to be reflected from the TR, the wave field in region 2 should be evanescent. An adequate plasma particle motion in that plasma region (region 2) is enforced. The corresponding attenuation rate of the Alfven wave field should be determined. The depth of the large-scale Alfven wave field penetration depends on the type of the wave field (electric of magnetic one), its orientation (either in the plane of incidence or perpendicular to it), whether the boundary is free or fixed, and the scales of changes of plasma parameters outside and inside the three plasma structures of the system under consideration: the upper chromosphere, TR and the lower corona. Our examination demonstrates that the depth of the field penetration of the shear Alfven wave field is close to the perpendicular wavenumber $k_{\perp}$ (19,21).

All previous analyses of kinetic Alfven waves have been focused on the strong Landau dissipation mechanisms. Large-scale responses due to the plasma-plasma boundary, like the transition region TR, have been omitted.

Third, the boundary conditions of Alfven wave fields are twofold. The first one refers to the electrodynamical point of view and is derived from Maxwell equations applied over the boundary (TR). The dynamical boundary conditions concern particle fluid motion and should consistently complement the electrodynamical boundary conditions. The fluid motion at the boundary depends on whether the boundary is fixed or free. The continuity of the bulk velocities (either fluid $\boldsymbol{u}$, or electron one $\boldsymbol{v}_e$) reads as

$$\boldsymbol{u}\,(1)\;(\boldsymbol{v}_e\,(1) = \boldsymbol{u}\,(2)\;(\boldsymbol{v}_e\,(2)).$$

under free boundary conditions. The dynamical boundary conditions also imply those for the continuity of the electric currents: $j_{\parallel,n}(1) = j_{\parallel,n}\,(2)$, which is in accordance with the continuity of the displacement field $\boldsymbol{D}$: $D_n(1) = D_n(2)$, i.e.

$$\varepsilon(1)E_n(1) = \varepsilon(2)\,E_n(2)$$

where $\varepsilon(i)$ denotes the dielectric function of region $i$ ($i$ =1,2), and $E_n$ (i) refers to the electric field components normal to the boundary $z = 0$.

The electric displacement field $\boldsymbol{D}$ is considered mainly in a kinetic aspect where a non-zero, time-varying electric field generally breaks the strict magnetic-kinetic equipartition, allowing energy to be converted and exchanged with plasma particles. The electric displacement field $\boldsymbol{D}$ represents a collective (large-scale or fluid-like) response. Its role in collective processes of mode conversion has not been discussed in the literature. Actually, a much stronger breach of the energy density equipartition in favor of electric energy density occurred through such collective (fluid-like) responses. Kinetic Alfven wave-ion-sound wave conversion is unavoidable even for large scales, homogeneous plasmas, and becomes intensified with the increase of the perpendicular wavenumber (the coupling parameter).

In idealized, low-frequency, incompressible MHD, Alfvén waves manifest perfect equipartition: the magnetic energy density equals the kinetic energy density, whereas the electric energy is negligible. Effects of the electric displacement field (***D***) arise whenever the ideal magnetohydrodynamic (MHD) approximation breaks down due to plasma-plasma inhomogeneity (boundaries), kinetic effects, high frequencies, finite-beta plasma environments, etc. Parallel electric field ($E_{\parallel}$) then emerges and changes the energy density equipartition of ideal Alfvén waves. Associated kinetic effects become most effective when the wave's perpendicular wavelength $k_{\perp}$ is comparable to the ion gyroradius. Then a significant parallel electric field is generated. The parallel electric field acts as a coupling factor that can transfer energy between the waves and charged particles (via Landau damping). Kinetic Alfvén waves also couple to ion sound modes, supporting electrostatic fluctuations that contribute to non-equipartition and enhanced wave dissipation.

Taking into account the electric displacement field ***D*** considered in the present study, the following cumulative effects on the KAW and ISW electric wave field transmission and reflection at the TR are highlighted:

i) Kinetic Alfven waves and ion sound waves are coupled modes even in uniform homogeneous magnetized plasmas;

ii) Compared to the ion sound spectrum in unmagnetized plasmas, the coupled ion sound wave spectrum decreases with the perpendicular wavenumber increase;

The novel results are:

iii) We assert the equivalence and/or compatibility of the commonly accepted rule for the reflection of Alfvén waves, the continuity condition for the particle motion (under free boundary condition), and the electric displacement field ***D***;

iv) The electric field (parallel and perpendicular to the background magnetic field $\boldsymbol{B}_{o}$) of the kinetic Alfven waves should increase its intensity (steeply) across the solar transition region. The amplitude increase rate is proportional to the ratio of the plasma densities of the upper chromosphere to that of the base of the corona, irrespective of the orientation of the magnetic field to the normal of the transition region;

v) The depth of the kinetic Alfven wave evanescent field zone beyond the reflection point is determined by the horizontal wave number;

vi) Given the enhanced electric fields at the base of the corona, a ponderomotive force is involved in the evanescent fields zone of the kinetic Alfven and possibly the ion-sound waves. This force is acting on plasma particles there, accelerating them upward to the corona;

vii) The plasma particle acceleration in the evanescent field zone does exhaust energy from the kinetic Alfven waves reflected back into the chromosphere. That energy is deposited at the base of the corona.

The proposed model of coupled kinetic Alfven-ion sound waves is adequate in the large-scale limit, when the wavelengths exceed the transition region thickness. The model assumes a simplified geometry for the transition region, treating it as an infinitely thin sheet, and smaller-scale processes occurring there are dismissed. The ambient magnetic field is locally uniform and oriented in an arbitrary direction; the plasma parameters outside (in the upper chromosphere and the base of the corona) are assumed locally homogeneous.
The more generalized conclusion is that the penetration of a large-scale Alfvén wave field through the transition region TP generates an amplification of the wave electric field as well as a zone of action of the ponderomotive force beyond the transition region, in the base of the solar corona. In this specific zone of ponderomotive force, plasma particles are accelerated synchronously upwards to the base of the corona. The kinetic energy acquired in that ponderomotive field zone can become comparable at least close to the thermal energy of the particles in the corona (of $10^2$ eV). Subsequent particle deceleration is expected, which suggests electromagnetic wave emissions possibly in the MHz range of radio frequencies.

In principle, the proposed model of large-scale Alfven wave energy conversion across the transition region fails to take into account small-scale processes occurring there.

Other conversion mechanisms proposed: non-linear wave interactions and turbulence, wave-wave conversions in inhomogeneous plasmas, enhanced Landau damping, etc., are important. They are of high efficiency, especially for much smaller spatial scales, comparable to the ion gyroradius, ion inertial length, and Debye length, etc., because the kinetic effects become dominant at these scales. All these small-scale processes are out of the scope of the present study.

## 5. Conclusion

The proposed mechanism of wave conversion induced by Alfvén wave reflection at the TR and possible acceleration of plasma particle in the Alfven wave evanescent field zone operates under large-scale conditions. The proposed mechanism of kinetic Alfven wave energy conversion is modelled within the two-fluid approximation. This kinetic Alfven wave conversion implies an initiation of electrostatic ion sound wave (ISW) with enhanced wave amplitudes and accelerated upward plasma particles. Both processes emerge in the Alfven wave evanescent field zone.

It is widely accepted that sources of high-intensity electric fields in the solar atmosphere are magnetic reconnection sites and flares of various scales, where plasma particles are accelerated, providing energy fluxes necessary to heat the corona and drive the solar wind. High-resolution observations of electric field signals from the transition region, in particular are desirable. The proposed Alfven wave mode energy conversion: steep enhancements of the Alfvén waves' electric field in the TR, and kinetic energy release in the evanescent Alfven wave field zone, implies an efficient Alfven wave energy supply to the base of the corona on large scales.
Overall, the up-to-date scarcity of electric field measurements remains the missing link for revealing how the solar corona/solar wind system is heated to millions of degrees. The hope is that present and future experiments, such as NASA's Parker Solar Probe and ESA's Solar Orbiter, designed to measure DC and fluctuating electric fields, would help discover and trace the electric fields of any sales back to the solar transition region TR, to understand the long-lasting coronal heating problem.

## APPENDIX

### Two-fluid model estimations of the transmitted kinetic Alfven waves

Consider the solar transition region TR as an interface between the upper chromosphere and the base of the solar corona. To understand the kinetic dynamics between the upper chromosphere and the base of the corona, the following considerations are taken into account: the two-fluid hydrodynamics equations, continuity, and momentum equations for electrons and ions are involved

$$\frac{\partial n_{e,i}}{\partial t} + \nabla.\left(n_0 v_{e,i}\right) = 0 \qquad \text{(A1)}$$

$$n_0 \frac{\partial n \boldsymbol{v}_{e,i}}{\partial t} = -q_{e,i}(\boldsymbol{E} + \boldsymbol{v}_{e,i} \times \boldsymbol{B}_0) - \nabla p_{e,i} \qquad \text{(A2)}$$

where index i (= e,i ) denotes electron and ion fluids, respectively. The background magnetic field $\boldsymbol{B}_0$ is for convenience assumed upward (perpendicular) to the TR. We look for kinetic Alfven and ion sound waves supported by a parallel electric field $E_{\parallel}$ coinciding with $E_z$. The Poisson equation for the electric field $\boldsymbol{E}$ is involved in completing the system of equations describing the particle dynamics in the system.

$$\nabla.\boldsymbol{E} = \frac{e}{\varepsilon_0}(n_i - n_e) \qquad \text{(A3)}$$

Divergences of (2) are taken and added to equations (1) and gathered together. Assuming the TR is an infinitely thin layer, the integration of (A4) in z (the height) is straightforward. Taking into account the continuity of the pressure forces across the TR, the following relationship is obtained:

$$n_{0,1}\left(1 + \frac{m_e}{m_i}\right) E_{\parallel,1} = n_{0,2}\left(1 + \frac{m_e}{m_i}\right) E_{\parallel,2} \qquad \text{(A4)}$$

together with the pressure force continuity condition at the boundary $z = 0$:

$$\gamma \frac{kT_{e,1}}{m_e} \frac{\partial \delta n_{e,1}}{\partial z} + \gamma \frac{kT_i}{m_i} \frac{\partial \delta n_{i,1}}{\partial z} = \gamma \frac{kT_{e,2}}{m_e} \frac{\partial \delta n_{e,2}}{\partial z} + \gamma \frac{kT_{i,2}}{m_i} \frac{\partial \delta n_{i,2}}{\partial z}. \qquad \text{(A5)}$$

For kinetic Alfven waves transmitting the TR, the above relationship yields the following ratio of the parallel wavenumbers $k_{\parallel,1}$(in the upper chromosphere) and $k_{\parallel,2}$ (in the base of the corona):

$$\frac{k_{\parallel,2}}{k_{\parallel,1}} = \frac{\frac{kT_{e,1}}{m_e}\delta n_{e,1} + \frac{kT_i}{m_i}\delta n_{i,1}}{\frac{kT_{e,2}}{m_e}\delta n_{e,2} + \frac{kT_{i,2}}{m_i}\delta n_{i,2}} \approx \frac{T_{e,1}}{T_{e,2}} \frac{\delta n_{e,1}}{\delta n_{e,2}} \cong \frac{E_{\parallel,1}}{E_{\parallel,2}}, \qquad \text{(A6)}$$

i.e.

$$\frac{k_{\parallel,2}}{k_{\parallel,1}} \ll 1,$$

suggesting that the transmitted kinetic Alfven waves should propagate almost horizontally and ultimately dissipate in the base of the corona.

Relations (A4-A6) are adequate in the two-fluid model examination only if the conditions

$$\omega^2 \ll k_{\parallel,1(2)}^2 \lambda_{De,1(2)}^2$$

is fulfilled: ω is the frequency of the kinetic Alfven wave, and $\lambda_{De}$ is the Debye length, assumed to hold in both regions 1 and 2.

**FIGURE CAPTIONS**

Fig. 1. Sketch of the upper chromosphere-transition region-the base of the corona system. The background magnetic field is oblique to the normal of the transition region boundary.

Fig. 2. Sketch of the zone of the Alfven wave evanescent field. The ponderomotive force zone is delineated.

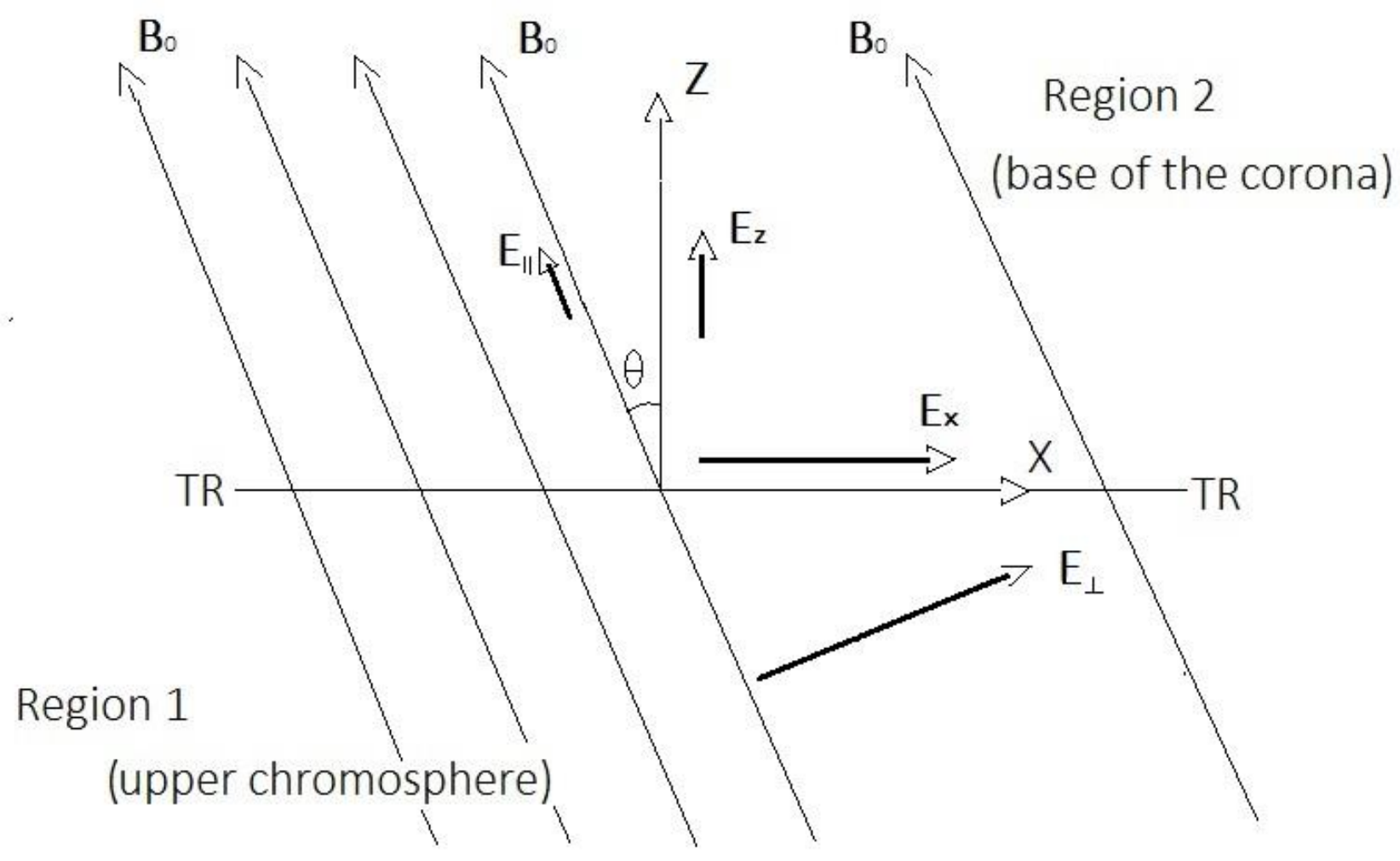

Figure 1

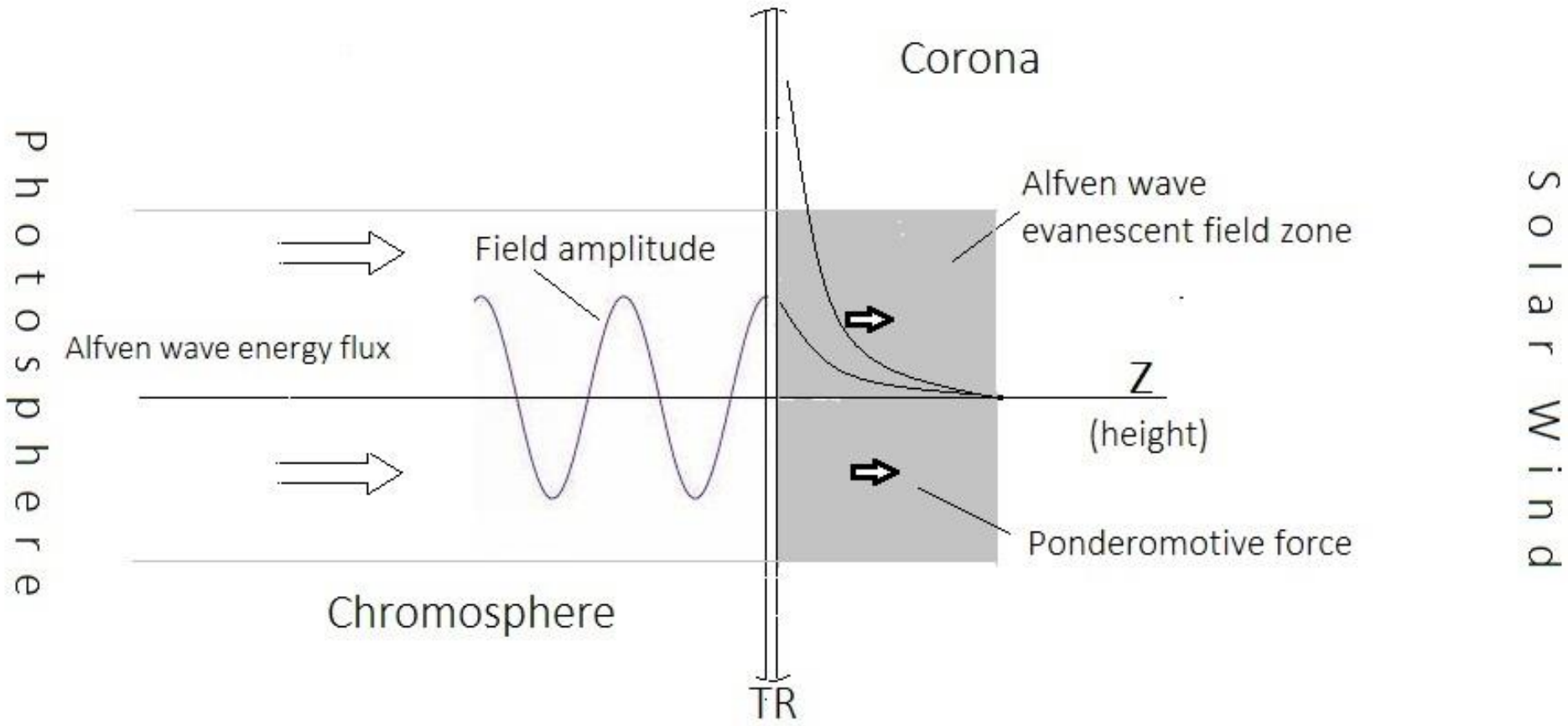

Figure 2